\documentclass[12pt]{article}
\usepackage{epsf}
\newcommand{\eqb}{\begin{equation}}
\newcommand{\eqe}{\end{equation}}
\newcommand{\dmb}{\begin{displaymath}}
\newcommand{\dme}{\end{displaymath}}
\newcommand{\pd}{\partial}
\newcommand{\ep}{\varepsilon}
\newcommand{\eab}{\begin{eqnarray}}
\newcommand{\eae}{\end{eqnarray}}

\newcommand{\e}{\mbox{e}}
\newcommand{\be}{\begin{equation}}
\newcommand{\ee}{\end{equation}}

\begin{document}
\begin{titlepage}
\begin{flushright}
TPI-MINN-00/29 \\
UMN-TH-1909/00   \\
June 2000
\end{flushright}
\vspace{0.6cm}
\begin{center}
\Large{{\bf BPS Saturated Domain Walls along a Compact Dimension}}
\vspace{1cm}

R. Hofmann and T. ter Veldhuis
\end{center}
\vspace{0.3cm}
\begin{center}
{\em Theoretical Physics Institute, Univ. of Minnesota, 
Minneapolis, MN 55455}
\end{center}
\vspace{0.5cm}
\begin{abstract}
Generalized Wess-Zumino models 
which admit topologically non-trivial 
BPS saturated configurations
along one compact, spatial dimension are 
investigated in 
various dimensions of space-time. We show that, in a representative model and
for sufficiently large circumference, there are 
BPS configurations along the compact dimension containing an arbitrary 
number of equidistant, well-separated domain walls. We analyze the spectrum
of the bosonic and fermionic light and massless modes that
are localized on these walls. 
The masses of the light modes are exponentially suppressed by the ratio
of the distance  between the walls and their width.
States that are initially localized
on one wall oscillate in time between all the walls. 
In (2+1) dimensions  the ``chirality'' of localized, massless fermions is 
determined. In the (1+1)-dimensional 
case we show how the mass of certain classically BPS saturated solitons
is lifted above the BPS bound by instanton tunneling.
\end{abstract} 
\vspace{4.0cm}
\begin{flushleft}
E-mail: 
rhofman@hep.umn.edu, 
veldhuis@hep.umn.edu
\end{flushleft}
\end{titlepage}

\section*{Introduction}

Recently, there have been various proposals entertaining
the idea that our $d=3+1$
dimensional world is located on a brane in a higher dimensional
space-time. In such scenarios, the standard model fields are
confined to the brane, whereas gravity permeates the bulk. The idea of 
compactification by 
localization of zero modes on defects of a higher dimensional 
field theory was 
first advocated in Ref.\,\cite{Shaposhnikov}.

In Refs.\cite{Dim1,Dim1b,Dim2,Dim3}, the hierarchy problem is addressed
through the notion of $n$ compact (large) extra 
dimensions of size $R\le 1\,{\rm mm}$. 
In this scenario the Planck scale $M_4$ of our world 
is not fundamental but can be derived from a 
higher dimensional scale $M_{4+n}$ of 
the order of the electroweak scale $M_W$  
by means of the following {\em power-law} conversion:
\eqb
M_4^2\sim M_{4+n}^{2+n}R^n\ .
\eqe
However, there is still a large 
hierarchy of the fundamental scales $M_{4+n},1/R$.   

In contrast, 
the authors of Ref.\,\cite{RandSund} 
suggest to solve the hierarchy problem in a rather different way. 
The set-up is a five-dimensional universe 
consisting of one three-brane constituting our world and another, 
hidden three-brane, which are separated by a (small) distance 
$\pi r_c$. Assuming a special, non-factorizable metric  
for the five dimensional space-time, there is practically no 
conversion between (fundamental) Planck scale $M_5$ and the 
Planck scale $M_4$ of our world. 
Scales $M$ belonging to the electroweak regime 
are given in terms of $M_5$ as
\eqb 
M=\e^{-2\pi k r_c} M_5\ ,
\eqe
where the fundamental scales $k,1/r_c$ 
have no large hierarchy ($kr_c\sim 50$).  

In addition, in Ref.\cite{DS2000} it was proposed
that the fermion hierarchy 
in the standard model may be generated by localizing 
different families on spatially separated walls  
along a compact dimension. Thereby, the localized light modes obtain their masses 
by coupling to a Higgs condensate which falls off exponentially away
from the wall on which the 
Higgs field is localized. 
 
The purpose of this paper is to study the idea of ``brane worlds'' in
a very simple supersymmetric context, namely Wess-Zumino models.
One and the same field is used to create domain walls (which play the
role of the branes) and the fermionic and bosonic light
modes localized on them. The latter  play the role of matter 
in the lower dimensional universes on the walls. 
We will assume the geometry of space-time as given, and we do not investigate 
the issue  of radius
stabilization, although the model we discuss in fact provides
a realization of ``brane-lattice-crystallization'', proposed
in Ref.\cite{Dim3} as a solution to this problem. The models we consider 
feature several distinct
topological sectors. We do not address the question why a particular sector 
is selected. 
Obviously, such a limited approach poses severe restrictions: there is no 
gravity,
no gauge fields, and we have to work in lower dimensions. The 
advantages are transparency, the ability to look at different issues
in isolation, and the potential to perform explicit calculations. We
will be able to address issues such as spontaneous supersymmetry breaking,
the chirality of the localized fermionic modes, and the spectrum of the
localized light modes. 

For concreteness, we choose to work with a specific model that
has the property that we desire; static configurations of well
separated domain walls in a compact space dimension. We stress that
it is just a toy model; the conclusions we draw should not, and do
not, depend on its specific details.

\subsection*{The model}
To be specific, 
we consider a generalized Wess-Zumino model in $d=3+1$
dimensions with one chiral superfield $\phi$.\footnote{We 
denote the superfield and its lowest component by the same
symbol $\phi$.} The  superpotential of this model is given by
\begin{equation}
\label{SUPOT}
W = \ln\,(\phi+1) + \ln\,(\phi-1)\ ,
\end{equation}
and the K\"{a}hler potential takes the canonical form. We also consider 
this model reduced to $d=2+1$ and 
$d=1+1$ dimensions.

Since the coordinate $z$ is compact, the field $\phi$ 
obeys periodic 
boundary conditions $\phi(x)=\phi(x+L)$, where $L$ is the circumference of the
compact dimension.
In general, static, stable solutions to the 
second-order equations of motion may or may not 
be BPS saturated. 
If a BPS 
saturated solution is selected as
the vacuum of the model, 
only 1/2 of the SUSY generators is spontaneously broken. 
We will focus mostly on the BPS saturated solutions, but static,
stable, non-BPS solutions also exist, and we will comment
on them in passing.


Since the presence of a non-vanishing
$(1,0)$ central charge $Z$ in the SUSY 
algebra is necessary for the 
existence of BPS domain walls, it was argued in 
Ref.\,\cite{Losev} that, in 
generalized Wess-Zumino models with a compact
dimension, the superpotential must be a multi-branch function and that
the manifold $M$ on which the  fields are defined must admit non-contractable 
cycles. The model with the superpotential $W$ of 
Eq.\,(\ref{SUPOT}) fulfills these criteria. The manifold $M$ in this
case is a plane with two punctures a the branch points $\phi=\pm1$.

As we also have occasion to consider the model 
in less than $d=3+1$ dimensions, and as
the scalar components of the
${\cal N}=1$ super multiplets in lower dimensions are real, it is convenient
to decompose the field $\phi$ according to 
$\phi=1/\sqrt{2}(\phi_1+\imath\phi_2)$ with real fields $\phi_1$ and $\phi_2$. 
By definition, the form of the scalar potential
is not altered by the procedure of dimensional reduction. 
The BPS equation and its solutions are the same
in each number of space-time dimensions that we consider. 
Therefore, we will first characterize all  
periodic solutions to the BPS equation; 
armed with this knowledge, we will subsequently investigate various 
implications in different dimensions. Of course, a solution along a
compact dimension is constrained by the requirement that
its period $\lambda$ fits
an integer $N$ times on the circumference $L$.

The scalar potential of the model 
is given by
\begin{equation}
V = \frac{dW}{d\phi} \frac{d\bar{W}}{d\bar{\phi}}=
 \left| \frac{1}{\phi +1} + \frac{1}{\phi-1}\right|^2.
\end{equation}
This potential is invariant under the transformations
$\phi_1 \rightarrow -\phi_1$ and
$\phi_2 \rightarrow -\phi_2$. It has a supersymmetric
vacuum, $V=0$ at $\phi=0$, and also a run-away vacuum at
$|\phi|\rightarrow \infty$. In the vacuum at the origin, the mass
of the field is $m=2$.
In addition, there are two saddle points at $\phi=\pm \imath$, and
two poles at $\phi = \pm 1$. As we will discuss in
detail below, the solutions to the BPS equation
either wind around one of these poles, or around both. 

At this point we briefly pause to discuss radiative corrections
to the classical BPS solutions and the spectrum of the localized
modes. Obviously, the model with the superpotential Eq.(\ref{SUPOT})
is not renormalizable in $d=3+1$ dimensions. Clearly, it
should be regarded as an effective theory, describing the low
energy dynamics of a fundamental theory below an energy scale $M$.
The field $\phi$ is supposedly the only relevant degree of freedom
below this scale. 

In fact, in order for an effective theory to be useful, it should be in
the weakly coupled regime. To that end, the superpotential in
Eq.(\ref{SUPOT}) can be slightly modified by the introduction of a
coupling constant $g$, 
\begin{equation}
\label{mod_SUPOT}
W = \ln\,(g \phi+1) + \ln\,(g \phi-1)\ ,
\end{equation}
so that for $g \ll 1$ the model is certainly weakly coupled.

The tension of the domain walls is
$T=2 \pi$, independent of $g$. The scale $T$ should be much smaller than 
the scale $M$ for consistency of the effective theory approach. Moreover,
the mass in the supersymmetric vacuum at the origin now
becomes $m=2 g^2$, and the mass scale $m_{pzm}$ of the pseudo-zero modes
localized on the wall is exponentially suppressed by the 
ratio of the distance between the walls and their width.
We thus find the folowing hierarchy of scales
\begin{equation}
M \gg T \gg m \gg m_{pzm}.
\end{equation}
It is the underlying fundamental theory that must create this hierarchy.
As a consequence,
it is possible to describe the dynamics in different energy
ranges by various effective theories.

Radiative corrections
to local quantities such as the energy density profile of the
domain wall can therefore be calculated in perturbation theory as
described in Refs.\cite{V,HS,Chib}. The scale $M$ forms a physical
ultra-violet cut-off in this calculation, and the size of the
corrections is controlled by $g$.

Moreover, the dynamics of the zero modes and pseudo zero modes
that are localized on the walls can be described by a weakly
coupled, $d=2+1$ dimensional, effective theory. In this case,
the scale $m$ forms the physical cut-off. Above this scale,
more massive modes become accessible and should be incorporated
in the theory. Radiative corrections can be calculated in 
a perturbative expansion and are controlled by powers of the
coupling constant $g$.

As radiative corrections are not the issue we investigate in this
paper, we will just take $g=1$ in what follows. Our results can
be trivially generalized for other values of $g$. In addition,
we will not speculate about the nature of an underlying fundamental theory
that could give rise to the superpotential in Eq.(\ref{mod_SUPOT}).

The paper can be outlined as follows: 
In Section \ref{sec0} we study and classify all periodic solutions
to the BPS equation. We show that there are BPS saturated configurations 
that contain well separated domain walls.
In Section \ref{sec1} we investigate such configurations
with a sequence of equidistant walls in $(3+1)$-dimensions.
Apart from the bosonic and fermionic zero modes associated with the
spontaneous breaking of translational invariance and half of
supersymmetry, there are other light modes with exponentially suppressed
masses. The sequence of domain walls can be considered as a crystal, and
the spectrum of the pseudo-zero modes is analyzed using tight binding
methods. We then study to what extend the domain walls can be considered
as parallel universes.
After integrating out the heavy modes, the  $(2+1)$-dimensional 
theory of the light modes has ${\cal N}=1$ supersymmetry. 
In the multi-wall background, 
light states that are localized 
on one particular wall are no longer mass eigenstates. This leads
to oscillation phenomena. Section \ref{sec2}  investigates the same model
in $(2+1)$ dimensions. The interesting new feature is that we are now
able to study the ``chirality'' (whether they are left or right moving)
of the fermionic zero modes localized on 
the domain walls (there are no chiral fermions in $d=2+1$ dimensions).
We show that in the BPS saturated background there are always one
left moving and one right moving fermion. We argue that this continues
to be the case even if the ${\cal N}=2$ supersymmetry is explicitly broken
to ${\cal N}=1$, in which case only one SUSY generator is
spontaneously broken. In Section \ref{sec3} we study the model in $d=1+1$ dimensions. What is
different here is that the domain walls with infinite mass in the higher
dimensions now become solitons with a finite mass. We show that in
particular topological sectors, and for sufficiently large circumference
of the compact dimension, the classical ground state is BPS saturated,
but non-perturbatively the ground state energy is lifted above the
BPS bound. This phenomenon was first observed in Ref. \cite{HLS}. 
The ground states are connected by instanton tunneling.
The tunneling probability and the energy shift of the ground state are 
obtained as a function of the circumference of the compact dimension. 
The Appendix contains technicalities concerning the fermionic zero modes.     

\section{Periodic solutions to the BPS equation \label{sec0}}

In this section we study periodic solution to the BPS equation,
\begin{equation}
\label{BPS-comp}
\frac{d\phi}{dx} = e^{\imath \delta} \frac{d\bar{W}}{d\bar{\phi}},
\end{equation}
where the phase $e^{\imath \delta}$ takes the value $\pm \imath$ for periodic
solutions.
The two possible choices for the phase correspond to whether the solutions wrap around
the poles clockwise or counter clockwise in the complex
$\phi$ plane. Solutions for one choice 
can be obtained from solutions with the opposite choice of the sign by
the transformation $x\rightarrow -x$, mapping walls into anti-walls and
vice versa. For definiteness, we will from here
on consider the BPS equation with $e^{i \delta}=+i$. It is well known that
solutions of the BPS equation have a ``constant of the motion''
$I={\rm Im}[e^{-i \delta } W]$. This constant of the motion can in principle be used
to mark the individual solutions. Here, instead of $I$, we use the related
quantity
\begin{equation}
k=e^{-2I}=\frac{1}{4}(\phi_1^2+2+\phi_2^2)^2 -2 \phi_1^2,
\label{conmot}
\end{equation}
to label the solutions, which takes a simpler form. 
The ``constant of the motion'' $k$ is positive semi-definite. It only 
vanishes at the poles $\phi=\pm 1$. Therefore, there are solutions for each 
value $k>0$. Since the BPS equation is a first order differential
equation, solutions, when viewed as trajectories in the 
complex $\phi$ plane, do not intersect each other or themselves,
except possibly at the supersymmetric vacuum $\phi=0$.

The target space is a plane with two punctures at the location of the poles
of the scalar potential. 
The trajectories of solutions to the BPS equation fall into
three homotopy classes of non-contractable 
cycles in the complex $\phi$ plane.\footnote{Here, we are actually
classifying the functions $f_k(s)=\phi_k(s \lambda(k))$ ($s \in [0,1]$), 
where $\lambda(k)$ is the wavelength of the solution labeled by $k$.}  
For any given value of $k$ in the range $0<k<1$ there
are two distinct solutions to the BPS equation. These two solutions are 
mapped onto each other by the transformation $\phi_1 \rightarrow -\phi_1$, which
is a symmetry of the scalar potential.
One solutions winds around  the pole at $\phi=1$, whereas the other winds
around  the pole at $\phi=-1$. We will denote these two homotopy classes as Class IA and
Class IB, respectively.
Solutions in a third homotopy class, which we refer to as Class II, are obtained for
$k>1$. For each value of $k>1$ there is only
one solution, which winds around both poles of the scalar potential and
is invariant under the transformation $\phi_1 \rightarrow -\phi_1$.

\begin{figure}
\epsfxsize=7cm
\centerline{\epsfbox{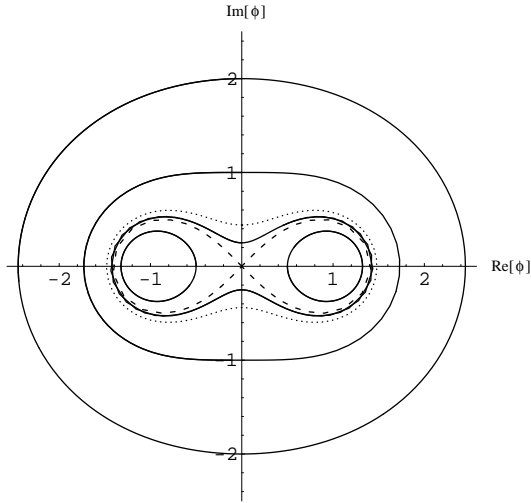}}
\caption{BPS solutions in the complex $\phi$ plane. The critical
solution is denoted by the dashed line. The 
dotted line indicates the Class II solution with minimum wavelength. Inside the
critical solution, a Class IA and a Class IB solution wind around their 
respective
poles. Also shown is a Class II solution passing through $\phi=1$, where
the scalar potential has a saddle point. Two other Class II solutions, one
with a value of $k$ near one (encompassing the
critical solution), the other with a larger value of $k$ (outermost cycle) 
are indicated.} 
\label{cycleP} 
\end{figure}

There is a critical solution for
$k=k_c=1$ which separates solutions in the three homotopy classes in the
complex $\phi$ plane. This critical solution contains
the point $\phi=0$, where the scalar potential has a minimum. 
The wavelength
of this solution is therefore infinite. It represents a domain wall that
interpolates between the same supersymmetric vacuum at $x \rightarrow -\infty$ and
$x \rightarrow +\infty$, winding either around the pole at $\phi=-1$ or the
pole at $\phi=+1$. We will refer to these walls as ``left'' and ``right'' walls
respectively.
Examples of solutions in each of the homotopy classes
are shown
in Fig.(\ref{cycleP}).
At $k=4$ the cycle passes through the two saddle points at
$\phi\pm 1$. For the cycles in Class II, the wavelength tends to infinity
both in the limit $k\downarrow 1$ and $k\rightarrow \infty$. 
In between, the wavelength reaches a minimum
which we numerically determined to take the value $\lambda_m=4.96$ for
$k_m=1.42$.

For very small values of $k$, $k\ll 1$, the cycles are small circles centered around
the poles in the complex $\phi$ plane. The wavelength of such solutions is approximately
$\lambda=\pi k/2$, and the energy density is approximately constant.
For large $k$, $k\gg 1$, the cycles are approximately large circles
centered around the origin. In this limit the wavelength is approximately given by 
$\lambda = \pi \sqrt{k}$, and again, the energy density is approximately
constant.
\begin{figure}
\epsfxsize=7cm
\centerline{\epsfbox{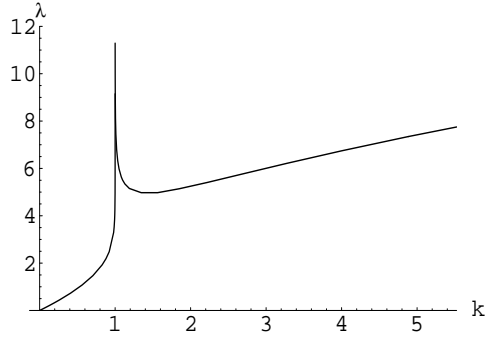}}
\caption{Wavelength $\lambda$ of the solutions as a function
of $k$. } 
\label{lamofk} 
\end{figure}

\subsection{Domain wall solutions}
\begin{figure}
\vspace{7cm}
\includegraphics{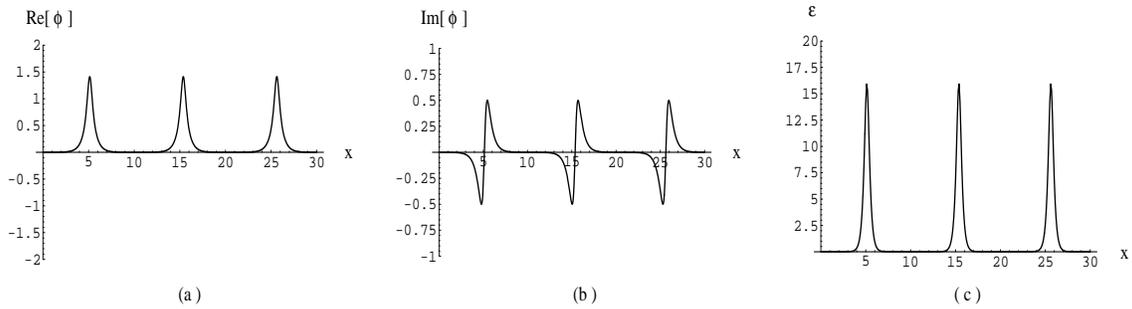}
\caption{Numerical solution to the BPS equation of Class IA subject to the initial condition 
${\rm Re}\left[\phi(0)\right]=0.0001$ and ${\rm Im}\left[\phi(0)\right]=0$. 
Depicted are (a) the real part of $\phi(x)$, 
(b) the imaginary part of $\phi(x)$, and (c) the energy density $\epsilon(x)$.} 
\label{numsolI} 
\end{figure}

\begin{figure}
\vspace{7cm}
\includegraphics{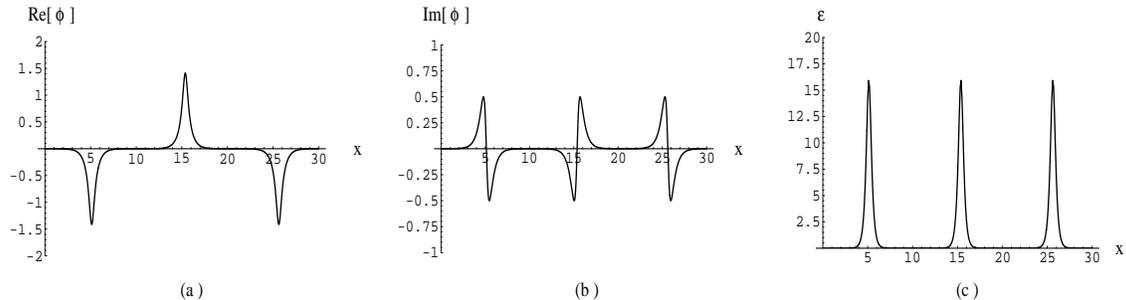}
\caption{Numerical solution to the BPS equation of Class II subject to the initial condition 
${\rm Re}\left[\phi(0)\right]=0$ and ${\rm Im}\left[\phi(0)\right]=0.0001$. 
Depicted are (a) the real part of $\phi(x)$, 
(b) the imaginary part of $\phi(x)$, and (c) the energy density $\epsilon(x)$.} 
\label{numsolII} 
\end{figure}

For $k$ close to the critical value $k_c=1$ the solutions 
exhibit well separated wall-like structures. To illustrate this,
we show in Fig.(\ref{numsolI}) a numerical solution $\phi(x)$ of Class IA 
subject to the initial condition ${\rm Re}\left[\phi(0)\right]=0.0001$ and 
${\rm Im}\left[\phi(0)\right]=0$. The value of the ``constant of the motion''
$k$ that is associated with this solution is $k\approx 1-10^{-8}$.
The corresponding 
energy-density $\ep(x)$ is also indicated. It is clear that if the trajectory
of the solution in the complex $\phi$ plane passes
the supersymmetric vacuum at the origin, the ratio of the width of the walls to distance
between them is
small since the force driving the trajectory away 
from the minimum is small. In Fig.(\ref{numsolII}) a numerical solution, 
belonging to Class II and subject to the initial condition 
${\rm Re}\left[\phi(0)\right]=0$ and 
${\rm Im}\left[\phi(0)\right]=0.0001$, is shown. 
The value of the ``constant of the motion''
$k$ that is associated with this solution is $k\approx 1+10^{-8}$.
For $k\uparrow1$ the solution consists of equally spaced ``right'' walls (Class IA) or
equally spaced ``left'' walls (Class IB), and for $k\downarrow 1$ the solution
consists of equally spaced, alternating ``left''  and ``right'' walls. As the
critical value $k_c=1$ is approached, the distance between the walls
increases, but the width and the shape of the walls converge.

\subsection{Approximate solutions in between the walls}
In between the walls, both the real and the imaginary part of $\phi$ are very small.
In this region the trajectory is approximately governed by the BPS equation 
linearized about $\phi=0$
\eab
\label{linBPSKP}
\frac{d\phi_1}{dx} & = & - 2 \phi_2\ ,\nonumber\\
\frac{d\phi_2}{dx} & = & - 2 \phi_1\ .
\eae
The general solution of Eq.\,(\ref{linBPSKP}) is
\eqb
\label{linsol}
\phi_1(x)=C_{+}e^{2 x}+C_{-}e^{-2x}\ ,\ \ \ 
\phi_2(x)=-C_{+}e^{2 x}+C_{-}e^{-2x}\ .
\eqe
For Class I cycles, with the initial condition $\phi_1(0)=u$ and $\phi_2(0)=0$,
the solution is
\begin{eqnarray}
\phi_1(x)=u \cosh 2 x\ , \nonumber \\
\phi_2(x)=-u \sinh 2 x\ .
\end{eqnarray}
The constant $u$, which is very small for well separated walls,
is related to the constant of motion $k$ of the solution
by $k=(1-u^2/2)^2$. The approximation breaks down when $x$ is
so large that the magnitude of the fields $\phi_1$ and $\phi_2$ becomes of the order one. 
This happens at the location of the walls. The wavelength of the solution is therefore
related to the constant $k$ by $\lambda=-1/2 \ln(1-k)$ for $k\uparrow 1$.

Similarly, the approximate solution for Class II solutions in between a 
``left'' wall
and a ``right'' wall is given by
\begin{eqnarray}
\phi_1(x) & = & -v \sinh 2 x\ , \nonumber \\
\phi_2(x) & = & v \cosh 2 x\ .
\end{eqnarray}
The constant
$v$, which also has to be very small for well separated walls, is related
to the constant of motion by $k=(1+v^2/2)^2$. From this approximate solution
it follows that the wavelength of the Class II solution is related to the
constant $k$  by $\lambda=-\ln(k-1)$ in the limit $k\downarrow 1$.

\subsection{BPS domain walls on a circle}

So far we have discussed generic periodic solutions to the BPS equation. Now
we want to determine what
BPS saturated configurations are possible on a compact space
with given circumference $L$. It is clear that we have to select only those
solutions in the total solution space for which the wavelength $\lambda$ fits
an integer $N$ times on the circumference $L$. The allowed values
of $k$ are thus the solutions of the equation $L=N \lambda(k)$, where 
the integer $N$ is a winding number that specifies how many times the
solution winds around the poles when $x$ varies between $0$ and $L$.
In Fig.(\ref{lamofk}) we show the period function $\lambda(k)$, which gives
the wavelength of a solution as a function of $k$. 

The simplest BPS configurations have winding number $N=1$; for any value of $L$
there is such a configuration of Class IA and Class IB, but only if $L>\lambda_m$
is there such a configuration of Class II. The tension $T\equiv 2\Delta W$ of the Class I configurations
is equal to $T=4 \pi$ while the tension of the Class II configuration
is equal to $T=8 \pi$. When $L$ is very large, the Class 
IA and IB configuration describe just one ``right'' wall and one ``left'' wall, respectively.
The Class II configuration contains one ``right'' wall and one ``left'' wall at
a distance of $L/2$.

Similar considerations apply to configurations with higher winding number $N$. 
The tension of the BPS configurations is proportional to $N$.
As long as $L$ is much larger than $N$ times the width of the walls, the configurations
contain well separated equidistant domain walls, $N$ ``right'' (``left'') walls in Class
IA (Class IB) configurations, and $N$ ``left'' walls alternating with $N$ ``right''
walls in Class II configurations.

\subsection{Non-BPS domain walls}

As explained in the previous section there is no Class II BPS saturated configuration
if the circumference $L$ of the compact dimension is smaller than $\lambda_m$.
However, there is a stable, static solution to the second order equations of
motion with the same topology. Such a solution can not be
BPS saturated. It is a non-contractable cycle, so
it can not be shrunk to a point. At the same time it has to remain at finite
distance of the origin in the complex $\phi$ plane because for fixed 
circumference $L$ the kinetic energy grows if the length of the trajectory
in the complex $\phi$ plane increases. 

Following the same argument, it is easy to see that there are stable,
static configurations that are not BPS saturated in all other topological
sectors. The simplest example is the homotopy class
of non-contractable cycles which wind first clockwise around one pole
and then counterclockwise around the other pole ("race-track" topology).
With obvious notation, we will refer to these topologies as 
$RL^{-1}$ and $LR^{-1}$.

Summarizing, in any non-trivial topological sector and for any
value of $L$ there is at least
one static, stable solution to the second order equation
of motion with minimal energy in that sector, a classical
ground state. In certain topological
sectors, those with the topologies $R^N$, $L^N$, $(R^{-1})^{N}$ and
$(L^{-1})^{N}$, the stable
solution is BPS saturated for any value of $L$, and only half
of supersymmetry is broken in the classical ground state. In other
topological sectors, those with the topologies $(RL)^N$ and
$(R^{-1}L^{-1})^N$,
there are two stable solutions that are BPS saturated if the circumference $L$
of the compact dimension is larger than a minimum value, $N \lambda_m$. 
If $L$ is smaller than the minimum value, then
there is still a static solution to the equation of motion, but it is
not BPS saturated.

\section{d=3+1; localized light modes \label{sec1}}

In this section we study the massless and light modes
in a BPS saturated background that takes the form of a sequence of equidistant
domain walls. We have discussed configurations of this type
in Section \ref{sec0}. As only half
of supersymmetry is spontaneously broken, the theory of the massless
and light modes in $d=2+1$ dimensions has ${\cal N}=1$ supersymmetry.
We will therefore  focus on the study of bosonic modes;
clearly, each bosonic mode is accompanied by a fermionic mode
that completes the supersymmetry multiplet.

There is an exact bosonic zero mode originating from the invariance of 
the action 
under translations along the compact dimension of the BPS saturated
background as a whole. 
We will now study the spectrum of the bosonic pseudo zero 
modes corresponding 
to independent translations of the domain walls\footnote{For 
large separation between the walls 
the action is nearly invariant
under independent translations of individual walls.}.

\subsection{Tight binding approach}
In oder to get a more quantitative picture of the bosonic pseudo zero modes,
we follow Ref.\,\cite{Chib} and 
expand the action in fluctuations of 
the fields around the classical, BPS saturated background configuration
up to quadratic order. 
In contrast to the case of a 
real wall background as it was investigated in Ref.\,\cite{Chib}, 
we have to keep track of the fact that here the BPS saturated wall 
$\phi_w$ is complex. 
Writing $\phi=\phi_w+\phi'$ and 
$\phi'=\frac{1}{\sqrt{2}}(\phi_1'+\imath \phi_2')$ with real fields $\phi_1'$ 
and $\phi_2'$, the following equations of motion are obtained
\eab
\label{eom-ab}
\left(\pd_\mu\pd^\mu+\bar{W}^{\prime\prime} W^{\prime\prime} +
\mbox{Re}\left[\bar{W}^\prime W^{\prime\prime\prime}\right] \right)\phi_1'
- \mbox{Im}\left[\bar{W}^{\prime} W^{\prime\prime\prime}\right] \phi_2'&=&0\ ,\nonumber\\ 
\left(\pd_\mu\pd^\mu+\bar{W}^{\prime\prime} W^{\prime\prime} -
\mbox{Re}\left[\bar{W}^\prime W^{\prime\prime\prime}\right] \right)\phi_2'-
\mbox{Im}\left[\bar{W}^{\prime} W^{\prime\prime\prime}\right] \phi_1'&=&0\ .
\eae
The primes denote derivatives
with respect to $\phi$ ($\bar{\phi}$) in the case of $W$ ($\bar{W}$) evaluated
at $\phi=\phi_w$ ($\bar{\phi}=\bar{\phi}_w$). For localized massless and
massive modes we make the following decomposition
\eqb
\label{ansatz}
\phi_1'=\chi_1(z) A(t,x,y)\ ,\ \ \ \phi_2'=\chi_2(z) A(t,x,y)\ ,
\eqe
where $\chi_i$ and $A$ are real fields. From the equation of motion, 
Eq.(\ref{eom-ab}),
we obtain the following equations of motion for these fields
\eab
\label{eom-albe}
-\frac{\partial_n \partial^n
A}{A}=\frac{\left(-\pd^2_{z}+\bar{W}^{\prime\prime} W^{\prime\prime}+
\mbox{Re}\left[\bar{W}^\prime W^{\prime\prime\prime}\right] \right)\chi_1 -
\mbox{Im}\left[\bar{W}^{\prime} W^{\prime\prime\prime}\right] \chi_2}{\chi_1}
\equiv m^2\ ,\nonumber\\ 
-\frac{\partial_n \partial^n
 A}{A}=\frac{\left(-\pd^2_{z}+ \bar{W}^{\prime\prime} W^{\prime\prime} -
 \mbox{Re}\left[\bar{W}^\prime W^{\prime\prime\prime}\right] \right)\chi_2-
\mbox{Im}\left[\bar{W}^{\prime} W^{\prime\prime\prime}\right] \chi_1}{\chi_2}
\equiv m^2\ , 
\eae
where $n=0,1,2$ is a Lorentz index in $d=2+1$ dimensions.
Hence, the field $A(t,x,y)$ 
has a mass 
$m$ in $d=2+1$ dimensions. The vector 
\begin{equation}
\vec{\chi}\equiv\left(\begin{array}{c}\chi_1\\ \chi_2\end{array}\right),
\end{equation}
satisfies the 
eigenvalue equation
\eqb
\label{eveq}
H\vec{\chi}=m^2\vec{\chi}\ ,  \label{evp}
\eqe
where the hermitian operator $H$ is defined as 
\eqb
\label{H}  
H=\left(\begin{array}{cc}
-\pd_{z}^2+\bar{W}^{\prime\prime} W^{\prime\prime}+
\mbox{Re}\left[\bar{W}^\prime W^{\prime\prime\prime}\right] &
-\mbox{Im}\left[\bar{W}^{\prime} W^{\prime\prime\prime}\right] \\
-\mbox{Im}\left[\bar{W}^{\prime} W^{\prime\prime\prime}\right]&
-\pd_{z}^2+\bar{W}^{\prime\prime} W^{\prime\prime}-
\mbox{Re}\left[\bar{W}^\prime W^{\prime\prime\prime}\right]
\end{array}\right)\ .
\eqe
This equation has the form of a one-dimensional Schr\"{o}dinger equation,
and, in analogy, we will refer to $H$ as the Hamiltonian and 
to $\vec{\chi}$ as the wave function.
Using the BPS equation, Eq.\,(\ref{BPS-comp}), it is not difficult to 
explicitly show that $H$ annihilates $\chi = \partial_z \phi_w$, which
corresponds to the 
Goldstone boson of spontaneously
broken overall translational invariance along the $\hat{z}$ direction.

Since we are not able to solve the eigenvalue problem in
Eq.(\ref{evp}) exactly, we resort to the tight binding approximation to
study the masses of the pseudo 
zero modes corresponding to independent translations of the individual
walls. 
For definiteness, we take the $R^N$ topological sector,  with 
the background of $N$ walls 
separated
by a distance $\lambda=L/N$.
Considerations for the $(RL)^N$ sectors are similar,
but slightly more complicated
due to the sequence of alternating walls and anti-walls. We will uncover the 
light bosonic modes
of the wall lattice by first considering a massless mode bound to a single 
wall in
isolation, just as one can find the electronic levels of a one dimensional
crystal from the energy levels of the individual atoms when they are far
apart and overlaps are small. 

Thus, we consider a single wall, $\phi_c^w$, along an infinite dimension. 
In this limit
$\chi^0\equiv\frac{1}{\sqrt{2\pi}}\partial_z \phi_c^w$ is a normalized eigen 
function with $m^2=0$.
In the case of $N$ such walls separated by an infinite distance,
there are $N$ localized (bound) wave functions with vanishing mass. 
However, when the 
walls are at a large, but finite, distance from each other, 
the $N$-fold degeneracy is lifted due to the small overlap of potentials 
and wave functions centered
at different wall sites, and a band structure emerges. 
We will now elucidate this band
structure, and we will estimate the width of the band.

To this end we decompose the operator $H^c=H^0 + V^c$, where
\begin{equation}
H^0 = \left(\begin{array}{cc}
-\partial_z^2+4 & 0 \\
0 & -\partial_z^2 +4
\end{array}\right),
\end{equation}
and
\begin{equation}
V^c = \left(\begin{array}{cc}
\bar{W}_c^{\prime\prime} W_c^{\prime\prime}- 4 +
\mbox{Re}\left[\bar{W}_c^\prime W_c^{\prime\prime\prime}\right] &
-\mbox{Im}\left[\bar{W}_c^{\prime} W_c^{\prime\prime\prime}\right]\\
-\mbox{Im}\left[\bar{W}_c^{\prime} W_c^{\prime\prime\prime}\right] & 
\bar{W}_c^{\prime\prime} W_c^{\prime\prime}- 4 -
\mbox{Re}\left[\bar{W}_c^\prime W_c^{\prime\prime\prime}\right]
\end{array}\right).
\end{equation}
Here the superscript $c$ indicates that the quantity is evaluated at
$\phi=\phi_c^w$. With this definition, $V^c$ vanishes exponentially fast away from the wall. We approximate
$H$ in the case of $N$ walls separated by a distance $\lambda$ as
\begin{equation}
H= H^0 + \sum_{n} V^c(z-n \lambda).
\end{equation}
This Hamiltonian commutes with the generator of the $R_N$ symmetry,
the translations over distances $n \lambda$.
The (pseudo zero mode) wave functions, 
the equivalent of Bloch waves, are linear
combinations of $\vec{\chi^0}$ centered at each of the $N$ wall sites. These
linear combinations transform
under irreducible representations of $R_N$. They take the form
\begin{equation}
\label{Bloch}
\vec{\chi}_k = \sum_{n} a_k\left[n\right] \vec{\chi}^0(z-n\lambda),
\end{equation}
where $\vec{a}_k$, with $k=1..N$, are the $N$ real, orthonormal eigenvectors of the 
$N \times N$ matrix
\begin{equation}
\left(
\begin{array}{ccccccccc}
2  & -1 &  0 & . & . & . & . & 0  & -1 \\
-1 &  2 & -1 & 0 & . & . & . & .  & 0 \\
0  & -1 &  2 &-1 & 0 & . & . & .  & 0 \\
.  &  . &  . & . & . & . & . & .  & . \\
0  &  . &  . & . & . & 0 & -1& 2  & -1 \\
-1 &  0 &  . & . & . & . & 0 & -1 & 2  
\end{array}
\right). \label{matrix}
\end{equation}
This matrix has a non-degenerate eigenvalue equal to zero for any
value of $N$, with eigenvector $\vec{a}_1=\frac{1}{\sqrt{N}}(1,1,..,1)$. For even
values of $N$ there is a second non-degenerate eigenvalue equal to $4$ with
eigen vector  $\vec{a}_2=\frac{1}{\sqrt{N}}(1,-1,1,..,-1,1)$. All other
eigenvalues are between $0$ and $4$ and are doubly degenerate. 
Therefore, the $N$ dimensional
representation of $R_N$ is reduced to $2$ one-dimensional 
and $N/2-1$ two-dimensional irreducible  representations in case
$N$ is even, and to $1$ one-dimensional 
and $(N-1)/2$ two-dimensional irreducible representations in case
$N$ is odd. 

As an aside, the eigenvectors and eigenvalues of the matrix in Eq.(\ref{matrix}) 
also yield the normal modes and frequencies of a closed linear chain of $N$ identical masses
and springs. In $d=3+1$ dimensions, the walls extend into 
two infinite dimensions and have therefore infinite mass. However, for solitons
in $d=1+1$ dimensions, which have finite energy, these vibrations are physical.
If
the distance between the solitons is large as compared to their width, 
these vibrations are very soft. 

Given the wave functions in Eq.(\ref{Bloch}), the approximate mass eigenvalues are then given by
\eqb
\label{varprob}
m_{TB}^2 = \frac{\int_0^L dz \vec \chi^T\,H\, \vec \chi}
{\int_0^L dz \vec \chi^T \vec \chi}.
\eqe
In fact, for the wave functions that transform under the two dimensional
representation of $R_N$, degenerate perturbation theory should be used.
Including only overlap integrals from neighboring sites, and
using the asymptotic form for $\vec{\chi}^c$ far away from the wall,
we estimate that the masses are exponentially suppressed,
\begin{equation}
m_{TB}^2=C e^{-2 \lambda},
\end{equation}
where $C$ is of the order one. We thus see that the $N$ degenerate vanishing masses 
in the limit where the walls are infinitely far apart evolve into a (mostly double degenerate)
band of masses when the distance between the walls becomes finite.

\subsection{Oscillations}
For an observer who can not resolve the circumference $L$ of the
compact dimension, the physics of the light modes is described by
an ${\cal N}=1$ supersymmetric theory in $d=2+1$ dimensions with
a spectrum as we discussed in the previous section. In such a world
there is a
hierarchy between the heavy modes, with masses comparable
to the wall tension, and the light modes, with masses that are 
exponentially suppressed by the ratio of the distance between the
walls and their width. It is interesting to note that in
topological sectors with multi-wall backgrounds that are not
BPS saturated the physics of the light modes is described by
a ${\cal N}=1$ model with supersymmetry breaking. The scale
of the breaking is equal to the scale of the light masses. It seems
that under such conditions there are no hierarchy or naturalness 
problems. 

Alternatively, an observer, who is confined to a particular wall and
can not resolve its width,
experiences
a different physical reality. Such an observer
does not have the means to do experiments in which the massive
bulk modes are excited. Still, due to the BPS saturation of the
background, his world has ${\cal N}=1$ supersymmetry. But he
will be perplexed by the
appearance and disappearance of particles in his world.  
From the ansatz of Eq.\,(\ref{Bloch}) 
it is clear that a state localized on 
a particular wall must be a 
superposition of the wave functions that represent the light modes. 
This in turn 
means that 
such a state is not a mass eigenstate. The implication 
is oscillation
of localized particles between the various walls with
frequencies corresponding to the mass differences. In the analogy of a
linear crystal, this would correspond to electrons hopping from atom to atom.
For example,  in the case of the $R^2$ sector with two walls, the approximate 
massless  mode is
\eqb 
\vec{\chi}_1(z)=\vec{\chi}^0(z)+\vec{\chi}^0(z-\lambda)\ ,\ \ 
\eqe
and there is one light mode, approximately given by
\eqb
\vec{\chi}_2(z)=\vec{\chi}^0(z)-\vec{\chi}^0(z-\lambda)\ .
\eqe
A state localized one of the walls 
is constructed by the superposition of the mass eigenstates,
\eqb
\vec{\chi}_0(z)\pm\vec{\chi}_1(z)\ .
\eqe

\section{d=2+1; chiral fermions \label{sec2}}

When the model is dimensionally reduced to $d=2+1$ dimensions, it has
${\cal N}=2$ supersymmetry. It has inherited four 
supersymmetry generators, whereas minimal supersymmetry in $d=2+1$ requires
only two generators.
The scalar potential may
be written as
\begin{equation}
V=\sum_{i=1}^2 \left(\frac{\partial{\cal W}}{\partial \phi_i}\right)^2.
\end{equation}
The object $\cal W$
plays the role of the 
superpotential in the dimensionally reduced model. 
It is related to the superpotential $W$ in $d=3+1$ dimensions by
\begin{equation}
{\cal W} = 2 {\rm Im} \left[ W\left(\frac{1}{\sqrt{2}}(\phi_1+\imath \phi_2)\right)\right]
=2 \arctan \frac{\phi_2}{\phi_1+\sqrt{2}} + 2 \arctan \frac{\phi_2}
{\phi_1-\sqrt{2}}\, .
\end{equation}
The extended supersymmetry manifests
itself in the fact that the superpotential $\cal W$ is harmonic.
In terms of the scalar components of the ${\cal N}=1$ superfields in
$d=2+1$ dimensions,
$\phi_1$ and $\phi_2$, the BPS equations are
\begin{eqnarray}
\label{BPSKP}
\frac{d\phi_1}{dx} & = & \alpha \frac{\partial {\cal W}}{\partial \phi_1} =
2 \alpha \left\{- \frac{\phi_2}{(\phi_1+\sqrt{2})^2 +\phi_2^2} -
\frac{\phi_2}{(\phi_1-\sqrt{2})^2 +\phi_2^2}\right\}, \nonumber \\
\frac{d\phi_2}{dx} & = & \alpha \frac{\partial {\cal W}}{\partial \phi_2} =
2 \alpha \left\{\frac{\phi_1+\sqrt{2}}{(\phi_1+\sqrt{2})^2 +\phi_2^2} + 
\frac{\phi_1-\sqrt{2}}{(\phi_1-\sqrt{2})^2 +\phi_2^2}\right\},
\end{eqnarray}
where $\alpha=\pm1$.
The interesting
new feature in $d=2+1$ dimensions is that now ``chiral'' fermions exist on the
$d=1+1$ dimensional worlds of the domain
walls. In $d=1+1$ dimensions there is no spin, but massless fermions can
be either left or right moving; this is what we will
allude to as ``chirality''. As outlined in the Appendix, in 
a BPS saturated background configuration there is always one left
moving and one right moving fermionic zero mode, corresponding to the
two spontaneously broken supersymmetry generators. Whether a mode
that is associated with a particular broken generator is left or
right moving depends 
on the sign $\alpha$ in the BPS equation that provides the background
configuration. 
In other words,
it is not important around {\em which} pole the background configuration winds 
in the complex $\phi$ plane, 
but whether it winds clockwise or counter clockwise.

\subsection{Wess-Zumino model}
In order to further investigate the issue of the ``chirality'' of massless 
fermions  on the wall, 
we  discuss in this section the  regular Wess-Zumino model reduced to 
$d=2+1$ dimensions.
It is well known
that this model allows for real, BPS saturated domain walls (kink or anti-kink). 
However, these solutions
do not exist in a compact dimension, and there are no BPS saturated multi-wall 
solutions (although {\em almost} static configurations with walls at
large distances do exist). For the issues we want to discuss here,
these differences are irrelevant.

Our analysis is similar to the discussion
of fermionic zero modes in Ref.\cite{SVV}, where the same model was 
investigated in 
$d=1+1$ dimensions.
The superpotential of the dimensionally reduced, regular Wess-Zumino model is
\begin{equation}
{\cal W} = \frac{1}{\sqrt{2}} \left( 2 \phi_1 - \frac{1}{3} \phi_1^3 + \phi_1 \phi_2^2 \right).
\end{equation}
There are two vacua, $\phi_1=\pm \sqrt{2}$ with $\phi_2=0$. The BPS saturated domain walls
connecting these vacua are determined by
\begin{eqnarray}
\frac{d\phi_1}{dy} & = & \alpha \frac{1}{\sqrt{2}} \left( 2 - \phi_1^2 + \phi_2^2 \right),
\nonumber \\
\frac{d\phi_2}{dy} & = & \alpha \frac{1}{\sqrt{2}} \left( 2 \phi_1 \phi_2 \right),
\end{eqnarray}
where $\alpha=\pm 1$. The solutions that interpolate
between the two vacua (the well-known kink or anti-kink, depending
on the sign $\alpha$), take the form
\begin{eqnarray}
\phi_1^w & = & \alpha \sqrt{2} \tanh (x-x_0)\nonumber \\
\phi_2^w & = & 0.
\end{eqnarray}
The equation of motion for the fermions in the background is
\begin{equation}
\imath \gamma^m\partial_m \psi_i -{\bf W}_{ij} \psi_j =0\ ,
\end{equation}
where the ``mass'' matrix ${\bf W}$ is given as
\begin{equation}
{\bf W} = \sqrt{2}
\left(
\begin{array}{cc}
-\phi_1^w & \phi_2^w \\
\phi_2^w & \phi_1^w
\end{array}
\right)\ .
\end{equation}
There are two massless fermions on the wall corresponding to 
Rebbi--Jackiw zero modes,
\begin{eqnarray}
\psi_1(x,y,t) & = & A(y) \zeta(x,t), \nonumber \\
\psi_2(x,y,t) & = & 0\ , \label{mode1}
\end{eqnarray}
with
\begin{eqnarray}
A(y)& = & e^{\alpha \int_{y_0}^{y} {\rm W}_{11}(y') dy'},\nonumber \\
\imath \gamma^2 \zeta & = & \alpha \zeta\ ,  \nonumber
\end{eqnarray}
and
\begin{eqnarray}
\psi_1(x,y,t) & = & 0, \nonumber \\
\psi_2(x,y,t) & = & B(y) \eta(x,t)\ , \label{mode2}
\end{eqnarray}
where
\begin{eqnarray}
B(y)& = & e^{-\alpha \int_{y_0}^{y} {\rm W}_{22}(y') dy'}, \nonumber \\
\imath \gamma^2 \eta & = & - \alpha \eta\ . \nonumber
\end{eqnarray}
Both $\zeta$ and $\eta$ satisfy the massless Dirac equation in
$d=1+1$ dimensions, $\imath \gamma^a\partial_a \zeta=0$, where $a=0,1$.
Note that $\gamma^2$ is the chiral projection matrix in $d=1+1$ dimensions, so
that one of the massless fermions is right moving and the other one is
left moving.
These massless fermions are the goldstino fermions associated with the
breaking of two out of four generators of the ${\cal N}=2$ supersymmetry, as discussed in 
the Appendix.

We proceed by addressing the fate of these massless fermions when the ${\cal N}=2$ supersymmetry
is explicitly broken to ${\cal N}=1$. In this case, the wall breaks
one of the remaining two supersymmetry generators, and therefore only
one Goldstone fermion exists. What happens to the second massless fermion
when the explicit supersymmetry breaking is switched on?
To answer this question we introduce explicit ${\cal N}=2$ breaking terms 
in the superpotential.

First, we insert a coupling constant $\lambda$
in front of the $\phi_1 \phi_2^2$ term in the superpotential, so that the
${\cal N}=2$ supersymmetry is explicitly broken unless $\lambda=1$. The vacuum expectation
values and the wall solution do not depend on $\lambda$, but ${\rm W}_{22}$ now
is equal to $\lambda \phi_1^w$. Even though there is only one
Goldstino, from the perspective of the Rebbi-Jackiw zero
modes in Eqs.\,(\ref{mode1},\ref{mode2}), it is clear that the second 
massless fermion continues to exist, even if $\lambda \ne 1$. When
$\lambda$ changes sign, the second massless fermion changes from right to left moving
(or vice versa, depending on the sign $\alpha$), and for $\lambda=0$ the second 
massless fermionic mode is not normalizable.

Next, we add a term $\frac{1}{2}\, \mu \phi_2^2$ to the superpotential, so that
the ${\cal N}=2$ supersymmetry is explicitly broken to ${\cal N}=1$ for $\mu \ne 0$. Again,
the vacuum expectation values and the wall do not depend on $\mu$, but
now ${\rm W}_{22}=\phi_1^w + \mu$. In this case
the second massless fermion  persists as long as $|\mu| < \sqrt{2}$. 
For $|\mu| > \sqrt{2}$ the second zero mode becomes non-normalizable.

It is clear that since the fermionic modes on the wall are chiral, they can only 
become massive in pairs, combining one left moving and with one right moving fermion. Individual
massless modes, either left or right moving, can only disappear when they become
non-normalizable. 

\subsection{Complex, multi-wall backgrounds}
In the case of a BPS saturated background with one
domain wall along a compact dimension, there is one massless scalar localized
on the wall corresponding to the spontaneously broken translational symmetry. 
Moreover, there is one left
moving and one right moving massless fermion, corresponding to the two 
spontaneously  broken supersymmetry generators (see the Appendix).
These massless modes combine to form a massless superfield in 
$d=1+1$ dimensions.

The supersymmetry multiplet in $d=1+1$ dimensions contains 
a real scalar and
a Majorana fermion. Therefore, in the case of $N$ walls, 
there will be a total of $N-1$ light superfields
($d=1+1$, ${\cal N}=1$ supersymmetry) 
with exponentially suppressed masses, and one massless superfield. 
In order to make states localized on a particular wall, 
states with different masses have
to be superimposed, just as in the $d=3+1$ dimensional case.

The explicit analysis of the Rebbi--Jackiw modes becomes
impossible if the fermion ``mass'' matrix in the background is not diagonal, 
as is
the case with the type of models we consider in this paper. 
But the results of the previous section must
still be valid. It is therefore expected that if the ${\cal N}=2$ 
supersymmetry
is explicitly broken to ${\cal N}=1$, there will still be a left and
a right moving
fermion on the wall, at least for a finite range
of the breaking parameters. Moreover, in a compact dimension, it
seems unlikely that fermionic zero modes can disappear by
becoming non-normalizable.

\section{d=1+1; solitons \label{sec3}}
The main new feature in $d=1+1$ dimensions is that the domain
walls, which have infinite energy in higher dimensions, now reduce
to solitons with a finite mass. Such solitons have a particle
interpretation, whereas the domain wall backgrounds in the higher
dimensions are interpreted as vacua.

All topological sectors in the $d=1+1$ dimensional theory, whether
they have a BPS saturated ground state or not, can be assigned
a pair of winding numbers $(N_l,N_r)$, where $N_l\in{\bf Z}$ is the 
winding number
around the ``left'' pole at $\phi=-1$ in Fig.(\ref{cycleP}), 
and $N_r\in{\bf Z}$ is
the winding number around the ``right'' pole at $\phi=+1$. 
According to our conventions
the winding number increases by one if a pole is encircled in the counter
clockwise direction. This classification of the sectors
according to the pair of winding numbers is not unique; for example,
the sectors $RLRL^{-1}$ and $R^2$ have the same winding numbers, $(0,2)$,
but they are topologically distinct.

The sectors $L^{N_l}$ and $R^{N_r}$ have a classical ground state in 
which only
half of supersymmetry is broken for any value of $L$. In the
$L^{N_l}$ sector this ground state
corresponds to the solution of the BPS equation which in the large $L$
limit contains $N_l$ ``left'' solitons if $N_l$ is positive, or $|N_l|$
anti ``left'' solitons if $N_l$ is negative, separated by a distance $L/|N_l|$.
The situation is analogous for the $R^{N_r}$ sectors.

For fixed $L > \lambda_m$, the sectors $(RL)^N$ have 
two degenerate classical ground states in which only
half of supersymmetry is broken. In the large $L$
limit, and if $N$ is positive, one of these ground states
corresponds to the solution of the BPS equation containing $N$ ``left''
solitons which alternate with $N$ ``right'' solitons, separated by a distance
$L/2N$. Similarly, if $N$ is negative we have the BPS solution with
$|N|$ ``anti-left'' solitons alternating with $|N|$ ``anti-right'' solitons
separated by a distance $L/2N$. The second, degenerated ground state corresponds to 
the BPS solution that winds around both poles at large distance, $N$ times in the
counterclockwise direction if $N$ is positive, and
$|N|$ times in the clockwise direction if $N$ is negative.
This solution to the BPS equation has approximately constant energy
density in the large $L$ limit. There is a tunneling transition with
finite action that connects the two ground states and the remaining
$1/2$ of supersymmetry is broken non-perturbatively.
For $L<N\lambda_m$, there is no BPS saturated ground
state in the sector $(RL)^N$ already at the classical level. 
There are static, stable
solutions to the second order equation of motion which
break supersymmetry completely at the classical level.

For the same reason,
supersymmetry is broken completely in the groundstate at the classical level
in all sectors other than the ones we discussed above.
In all sectors, the ground state energy tends to $2\pi(|N_l|+|N_r|)$ in the
large $L$ limit. The theory has two types of
solitons, ``left'' and ``right'', and two topologically conserved quantum numbers,
$N_l$ and $N_r$. The solitons 
interact through a short range force, 
which decreases exponentially with distance. ``Left'' and ``right'' solitons
repel each other, and so do ``anti-left'' and ``anti-right'' solitons. ``Left''
and ``anti-left'' solitons on the other hand attract each other.

\subsection{Semi-classical calculation of the tunneling amplitude}
\begin{figure}
\epsfxsize=7cm
\centerline{\epsfbox{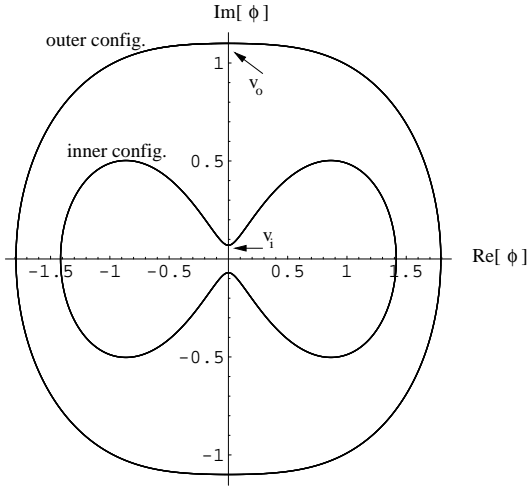}}
\vskip5mm
\caption{The initial (outer) and final (inner) configurations in the complex $\phi$ plane.
Both configurations are solutions to the
BPS equation with the same wavelength $\lambda$.
The instanton
interpolates smoothly between the initial and final configuration with finite action.} 
\label{instanton1} 
\end{figure}

In this section we  calculate the amount by which the
ground state energy is lifted above the BPS bound by the tunneling transition
in the sector $(RL)^N$, for $N=1$ and $L>\lambda_m$. We choose $N=1$ for 
definiteness; the tunneling
action scales linearly with $|N|$, and therefore the result can be trivially
generalized for any value of $N$.
The two ground states correspond to two solutions of the BPS equation 
in Class II, 
one with $k<k_m$ and one with $k>k_m$, as shown in Fig.(\ref{instanton1}). 
At the non-perturbative 
level the corresponding 
solitons are not longer BPS saturated because a tunneling process
mixes them and lifts their mass above the BPS bound. This phenomenon was first
observed in Ref.\cite{HLS}. In Ref.\cite{BSV} it was noted that the calculation
of the semi-classical instanton action is equivalent to the calculation
of the energy of domain wall junctions in $d=2+1$ dimensions. In the
same work, the general calculational framework
of the action was established and illustrated for a specific model. 

\begin{figure}
\epsfxsize=7cm
\centerline{\epsfbox{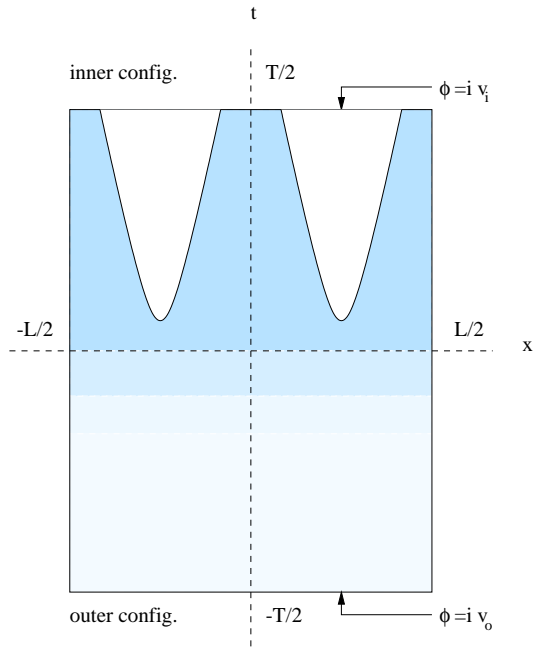}}
\caption{
Boundary conditions  in the $x$-$t$ plane. 
The initial and final configuration at the bottom ($-T/2$) and top ($T/2$) of
the rectangle are traced counter-clockwise in the
complex $\phi$ plane (see Fig.(\ref{instanton1})). Periodic boundary conditions apply to the 
vertical edges. The shading schematically indicates the energy density, with
the darker areas indicating higher energy density.} 
\label{instanton2} 
\end{figure}

We
assume that the instanton in the imaginary time saturates the BPS bound in the
model we consider; there is no obvious reason why this would not
be the case. If the instanton
is BPS saturated, it satisfies the equation
\begin{equation}
\frac{\partial \phi}{\partial \zeta} = \frac{1}{2} e^{i \delta}
\frac{d\bar{W}}{d\bar{\phi}},
\end{equation}
where $\zeta=x+ i t$, and $t$ is the euclidean time. The field $\phi$
depends both on $\zeta$ and $\bar{\zeta}$. The phase $\delta$ has to be
chosen so that it is consistent with the boundary conditions that are imposed
on the solution (see below).
The tunneling action takes the form
\begin{eqnarray}
A & = & 2 \int dx dt \left\{ \frac{\partial \phi}{\partial \zeta}
\frac{\partial \bar{\phi}}{\partial \bar{\zeta}}
+ \frac{\partial \phi}{\partial \bar{\zeta}} \frac{\partial \bar{\phi}}{\partial \zeta} \right\}
\nonumber \\
 & & + \int dxdt  
\frac{dW}{d\phi}
\frac{d\bar{W}}{d\bar{\phi}}.
\end{eqnarray}  
In Figs.(\ref{instanton1},
\ref{instanton2}) we
have indicated the boundary conditions that are imposed on the instanton configuration. 
At $t=-T/2$ the instanton field $\phi(x,t)$ takes
the form of the ``outer'' configuration (see Fig.(\ref{instanton1})) and
at $t=T/2$ it takes the form of the ``inner'' configuration. We always
have in mind the limit $T \rightarrow \infty$. Periodic boundary
conditions apply to the vertical edges in Fig.(\ref{instanton2}).
We have chosen the 
solutions to wind counter-clockwise in the $\phi_1$-$\phi_2$ plane when followed
in the positive $\hat{x}$ direction, which corresponds to $\delta=\pi/2$ in the
BPS equation. The initial and final configuration are positioned such that
$\phi(L/2,-T/4)=i v_0$ and $\phi(L/2,T/4)=i v_i$.
Note that in order for the instanton configuration to satisfy the
BPS equation, it has to connect the solution with the larger value of $k$ as
the initial configuration with the solution with the smaller value of $k$ as the
final configuration. The anti-instanton that interpolates in the reverse
situation satisfies an anti-BPS equation. The action can now be written as
\begin{equation}
A_{BPS} = \int dx dt\, \left[ 2 \vec{\nabla} \cdot \vec{S} -(\vec{\nabla}
\times \vec{a})_z\right],
\end{equation}
where
\begin{equation}
\vec{S} = \left(
\begin{array}{c}
\cos \delta\, {\rm Re}\left[W\right] + \sin \delta\, {\rm Im}\left[W\right] \\
\cos \delta\, {\rm Im}\left[W\right] - \sin \delta\, {\rm Re}\left[W\right] \\
\end{array} \right),
\end{equation}
and
\begin{equation}
\vec{a} = \left(
\begin{array}{c}
{\rm Im} \left[ \phi \frac{\partial \bar{\phi}}{\partial x} \right] \\
{\rm Im} \left[ \phi \frac{\partial \bar{\phi}}{\partial t} \right] \\
\end{array} \right).
\end{equation}
With the application of Gauss' and Stoke's theorems the surface
integrals can be converted into line integrals over the boundaries of the
surface,
\begin{equation}
A_{BPS} = 2 \oint \vec{S}\cdot  d\vec{n} - \oint \vec{a}\cdot d\vec{x},
\end{equation}
where the boundaries are the edges of the rectangle in Fig.(\ref{instanton2}).
Due to the periodic boundary conditions only the integrals of $\vec{S}$ contribute
over the vertical edges of the contour, as the superpotential is multi-valued.
Turning now to the explicit calculation of the tunneling action,
there are three non-vanishing contributions. First, consider
\begin{eqnarray}
A_{BPS}^0 & = & 2 \int_{-T/2}^{T/2} \left\{ {\rm Im} \left[ W \right]_{x=\frac{L}{2}}
- {\rm Im} \left[ W \right]_{x=-\frac{L}{2}} \right\} dt, \nonumber \\
 & = & 8 \pi T.
\end{eqnarray}
This contribution is equal to the classical ground state energy times the elapsed
time $T$. It is equal to the action in case there is no tunneling, when the
initial and final configuration are the same. We define the tunneling
probability by subtracting this trivial contribution in the exponent. 
Next, consider
\begin{eqnarray}
A_{BPS}^1 & = & 2 \int_{-L/2}^{L/2} \left\{ {\rm Re} \left[ W \right]_{t=-\frac{T}{2}}
- {\rm Re} \left[ W \right]_{t=\frac{T}{2}} \right\} dx \nonumber \\
 & = & \ln \left(\frac{k_0}{k_i}\right) L,
\end{eqnarray}
where $k_o$ and $k_i$ are the two solutions to the equation $\lambda(k)=L$ with
$k>k_c$. We define
$k_o$ to be the larger of the two. Note that
the integrands are ``constants of motion'' for the solutions to the
BPS equation, and therefore
the integrals can be expressed in $k_o$ and $k_i$, respectively. In the limit
$\lambda=L\gg \lambda_m$, $k_i$ approaches $1$ from above, and $k_o$ is
related to $L$ by $L= \pi \sqrt{k}$. In this limit, the second contribution
to the tunneling action is therefore $A_{BPS}^1=2 L \ln L/\pi$. Finally, we
calculate
\begin{eqnarray}
A_{BPS}^2 & = & \int \left\{ {\rm Im} \left[ \phi \frac{\partial \bar{\phi}}{\partial x}
\right]_{t=-\frac{T}{2}} -
 {\rm Im} \left[ \phi \frac{\partial \bar{\phi}}{\partial x}
\right]_{t=\frac{T}{2}} \right\} dx \nonumber \\
 & = & -S(k_o) + S(k_i),
\end{eqnarray}
where $S(k)$ is the area enclosed in the $\phi_1$-$\phi_2$ plane by the solution
to the BPS equation marked by
$k$. $S(k)$ is a monotonically increasing function of $k$. In the
limit $k\downarrow k_c=1$, $S(k)$ approaches $S(1)=4$ from above. 
For large values of $k$ on the other hand $S(k)=2 \pi \sqrt{k}$. Therefore, in the
limit $\lambda=L \gg \lambda_m$, the third contribution to the tunneling action is
approximately $A_{BPS}^2=-2 \pi \sqrt{k_o} + 1$, or, in terms of $L$, 
$A_{BPS}^2 = - 2 L + 1$. 

We have calculated the tunneling action solely in terms of quantities related
to the boundary configurations. An explicit interpolating instanton 
configuration is not needed, just the existence of a BPS saturated
instanton configuration is sufficient.
In the large $L$ limit, the semi-classical approximation is valid, 
and the ground state energy is lifted above the BPS bound by an amount
\begin{equation}
\Delta E \propto 8\pi e^{2 A_{BPS}} = 8 \pi\, e^{-4 L \ln L},
\end{equation}
to leading order in $L$. The solution to the BPS equation that
connects the vacuum at $\phi=0$ to the runaway vacuum at
$\phi \rightarrow i \infty$ (this is a solution that is not periodic, 
obtained with $\delta=0$) has
infinite tension. The circumference $L$ in a sense acts like an
infra-red regulator. For this reason, the tunneling action is not
linear in $L$, but proportional to $L \ln L$. We also note
that in this model the instanton mixes a state containing two 
localized solitons
with a state that corresponds to a flat energy density. This is in
contrast with the model considered in Ref.\cite{BSV}, where the
mixing is between two states with different solitons.

\section*{Summary}
We have constructed a toy model to study some issues related to 
the idea of ``brane worlds'' in a field theoretical context. This toy model
is a simple Wess-Zumino model with only one chiral superfield in
$d=3+1$ dimensions. Different aspects were studied in various
dimensions.

We showed that it is possible  to have BPS saturated configurations with
multiple, well separated domain walls in a compact
space dimension. Generically, in such a background
there are light modes with masses exponentially suppressed by
the ratio of the distance between the walls and their width.
These modes form the ``matter'' of the ``brane world'', wheras
the domain walls play the role of 
branes in our simple picture. We also described
how particles can oscillate between walls. 
We illustrated these issues in our specific model.  

We also found that the ``chiral'' fermions localized
on $d=1+1$ dimensional BPS saturated walls embedded in a $d=2+1$
dimensional space-time always come in pairs; one is left moving, and one 
is right moving.
 
Finally, we discussed the toy model in $d=1+1$ dimensions. The domain
walls then reduce to solitons with a finite mass. We showed that
some classically BPS saturated states are lifted non-perturbatively
by instanton tunneling. 

\section*{Acknowledgements}

The authors would like to thank M. Shifman for valuable discussions. 
T.t.V also acknowledges useful conversations with A. Losev and A. Vainshtein. 
The work of R.H. was supported by a postdoctoral fellowship 
of Deutscher Akademischer Austauschdienst (DAAD), the work of T.t.V. by
the Department of Energy under Grant No. DE-FG02-94ER40823.

\section*{Appendix A}

In this Appendix we establish our notation and discuss the zero modes
in a BPS saturated background.
\subsection*{d=3+1 dimensions}
We use the conventions outlined in \cite{Chib}. The signature of the metric $g^{\mu\nu}$
is $(+,-,-,-)$.
The component Lagrangian for an ${\cal N}=1$ Wess-Zumino model in $d=3+1$ dimensions
with canonical K\"{a}hler potential and superpotential $W$ is
\begin{eqnarray}
\cal{L} & = & \partial_{\mu} \bar{\phi} \partial^{\mu} \phi + \imath \psi^{\alpha}
\sigma_{\alpha \dot{\alpha}}^{\mu} \partial_{\mu} \bar{\psi}^{\dot{\alpha}}
+ F^{\dagger} F \nonumber \\
 & & + F \frac{d W}{d \phi} - \frac{1}{2} \frac{d^2 W}{d\phi^2}
\psi^{\alpha} \psi_{\alpha} \nonumber \\
 & & + \bar{F} \frac{d\bar{W}}{d\bar{\phi}}
-\frac{1}{2} \frac{d^2 \bar{W}}{d^2\bar{\phi}} \bar{\psi}_{\dot{\alpha}}
\bar{\psi}^{\dot{\alpha}}.
\end{eqnarray}
The component fields transform under the supersymmetry transformations as
\begin{eqnarray}
\delta \phi & = & \zeta^{\alpha} \psi_{\alpha}, \nonumber \\
\delta \psi_{\alpha} & = & \imath \sigma_{\alpha \dot{\alpha}}^{\mu}
\bar{\zeta}^{\dot{\alpha}} \partial_{\mu} \phi + \zeta_{\alpha} F, \nonumber \\
\delta F & = & \imath \bar{\zeta}_{\dot{\alpha}} \bar{\sigma}^{\mu \dot{\alpha} \alpha}
\partial_{\mu} \psi_{\alpha},
\end{eqnarray}
where the transformation parameter $\zeta$ is the two component complex Grassmann 
variable.
The auxiliary field $F$ can be eliminated by the use of its equation
of motion $\bar{F}=-\frac{d\bar{W}}{d\bar{\phi}}$ to give the Lagrangian
density
\begin{eqnarray}
\cal{L} & = & \partial_{\mu} \bar{\phi} \partial^{\mu} \phi + \imath \psi^{\alpha}
\sigma_{\alpha \dot{\alpha}}^{\mu} \partial_{\mu} \bar{\psi}^{\dot{\alpha}}
-\frac{dW}{d\phi} \frac{d\bar{W}}{d\bar{\phi}} \nonumber \\
 & & - \frac{1}{2} \frac{d^2 W}{d\phi^2} \psi^{\alpha} \psi_{\alpha}
-\frac{1}{2} \frac{d^2 \bar{W}}{d\bar{\phi}^2} \bar{\psi}_{\dot{\alpha}}
\bar{\psi}^{\dot{\alpha}}.
\end{eqnarray}
The BPS equation for domain walls reads
\begin{equation}
\frac{d\phi}{dz}= e^{\imath \delta} \frac{d\bar{W}}{d\bar{\phi}},
\end{equation}
where in a non-compact direction the phase $\delta$ is determined from the
equation
\begin{equation}
{\rm Im} \left[e^{-\imath \delta} W(\phi(z=-\infty)\right] =
{\rm Im} \left[e^{-\imath \delta} W(\phi(z=+\infty)\right].
\end{equation}
BPS domain walls in a compact dimension are only possible if the superpotential 
is not single valued. In the models we consider
$e^{\imath \delta}\pm i$. 

Let us now expand the Lagrangian of the model about the 
non-trivial solution $\phi_w$ of the BPS equation. Setting $\phi=\phi_w+\phi'$ 
and expanding up to
quadratic terms in the fluctuations $\phi'$, we obtain
\begin{eqnarray}
{\cal L} & = & \partial_{\mu} \bar{\phi} \partial^{\mu} \phi + 
\imath \psi^{\alpha} \sigma_{\alpha\dot{\alpha}}^{\mu} \partial_{\mu}
\bar{\psi}^{\dot{\alpha}} \nonumber \\
 & & - \frac{1}{2}\left\{  W'''\bar{W}'\phi'^2 +   W'\bar{W}'''\bar{\phi}'^2 
+2  W''\bar{W}''\phi'\bar{\phi}' 
+  W'' \psi^{\alpha} \psi_{\alpha}
+ \bar{W}'' \bar{\psi}_{\dot{\alpha}} \bar{\psi}^{\dot{\alpha}} \right\},
\end{eqnarray}
where the primes on $W$ ($\bar{W}$) indicate derivatives with respect to
$\phi$ ($\bar{\phi}$) evaluated at $\phi=\phi_w$ ($\bar{\phi}=\bar{\phi}_w$).
The equation of motion for the scalar in the wall background is
\begin{equation}
\partial^2 \phi' + \left\{W'\bar{W}'''\bar{\phi}' + W''\bar{W}''\phi'\right\}=0,
\end{equation}
and the equation of motion for the fermion in the wall background is
\begin{equation}
\imath \sigma_{\alpha\dot{\alpha}}^{\mu} \partial_{\mu} \bar{\psi}^{\dot{\alpha}}
-W'' \psi_{\alpha} = 0.
\end{equation}
Half of the supersymmetry generators is broken by the wall solution. In order
to determine which generators are unbroken, we act with the SUSY generators on the wall
and find
\begin{equation}
\delta \psi_{\alpha} = - \frac{1}{\alpha} \frac{d\phi_w}{dz} \left(
\zeta_{\alpha} + \imath \alpha \sigma_{\alpha\dot{\alpha}}^{3} \bar{\zeta}^{\dot{\alpha}}
\right),
\end{equation}
where we have used the BPS equation, and $\alpha=\pm 1$. Transformations, for which the 
parameter satisfies $\zeta_{\alpha}= \imath \alpha \sigma_{\alpha
\dot{\alpha}}^3 \bar{\zeta}^{\dot{\alpha}}$, correspond to unbroken symmetries.
The massless fermions localized on the wall  
which are related to the broken SUSY generators take the form
\begin{equation}
\psi_{\alpha}(x,y,z,t) = - \frac{d\phi_w}{dz}(z) \zeta_{\alpha}(x,y,t),
\end{equation}
where $\zeta_{\alpha}(x,y,t)$ is constrained by
\begin{equation}
\zeta_{\alpha} = + \imath \alpha \sigma_{\alpha\dot{\alpha}}^3 
\bar{\zeta}^{\dot{\alpha}}.\label{constraint}
\end{equation}
It can then be verified that $\psi_{\alpha}$ satisfies the fermionic equation of motion
in the domain wall background if $\zeta_{\alpha}$ satisfies the equation
\begin{equation}
\imath \sigma_{\alpha\dot{\alpha}}^{m} \bar{\sigma}^{3\dot{\alpha}\beta}
\partial_m\zeta_{\beta}=0,
\end{equation}
where $m$ takes the values $0,1,2$. This is exactly the Dirac equation
in $d=2+1$ dimensions,
\begin{equation}
\imath \gamma^m \partial_m \zeta = 0,
\end{equation}
with the following representation of the $\gamma$ matrices; $\gamma^0=\sigma^3$,
$\gamma^1=- \imath \sigma^2$ and $\gamma^2=\imath \sigma_1$.
In this basis the charge conjugation matrix, defined by $C^T=-C$ and $C\gamma^{\mu}C^{-1}=
-(\gamma^{\mu})^T$, takes the form $C=\pm \imath \sigma^2$. The Majorana constraint reads
\begin{equation}
\zeta^T C = \imath \zeta^{\dagger} \gamma^0.
\end{equation}
This is identical to the constraint in Eq.(\ref{constraint}). We thus
find a massless Majorana fermion in the $d=2+1$ dimensional world on the wall. Together with
the massless real scalar, this fermion forms a SUSY multiplet in the lower
dimensional theory.

\subsection*{d=2+1 dimensions}
\subsubsection*{${\cal N}=1$ SUSY}

The Lagrangian of the model with ${\cal N}=1$ supersymmetry and one superfield
with canonical kinetic terms and superpotential ${\cal W}$ takes the form
\begin{eqnarray}
\cal{L} & = & \frac{1}{2} \left\{ \partial_m \phi \partial^m \phi + 
\imath \bar{\psi} \gamma^{m} \partial_{m} \psi + F^2 \right. \nonumber \\
 & & \left. +2 F \frac{\partial{\cal W}}{\partial\phi} 
-\frac{\partial^2{\cal W}}{\partial\phi^2} \bar{\psi} \psi \right\}.
\end{eqnarray}
We choose the Majorana basis for the $\gamma$ matrices, $\gamma^0=\sigma^2$, 
$\gamma^1=\imath \sigma^3$ and $\gamma^2=\imath \sigma^1$. Here $\phi$ is
a real scalar field and $\psi$ is a real two component spinor, with the
Dirac conjugate defined by 
$\bar{\psi}=\psi^T \gamma^0$. The supersymmetry transformations take
the form 
\begin{equation}
\begin{array}{l}
\delta \phi =  \bar{\zeta} \psi\, ,  \\
\delta \psi =  - \imath \partial_{m} \phi \gamma^{m} \zeta + F\zeta\, , \\
\delta F = - \imath \bar{\zeta}\gamma^{m}\partial_{m} \psi\, ,
\end{array}
\end{equation}
where the transformation parameter $\zeta$ is a two component real Grassmann variable.
The BPS equation is
\begin{equation}
\label{BPS-3}
\frac{d\phi}{dy}= \alpha \frac{d{\cal W}}{d\phi},
\end{equation}
where $\alpha=\pm 1$.

Expanding the Lagrangian about the non-trivial solution $\phi_w$ 
to Eq.\,(\ref{BPS-3}) up to quadratic terms in the fluctuations, we obtain
\begin{equation}
{\cal L} = \frac{1}{2} \left\{
\partial_m \phi' \partial^m\phi' +  \imath \bar{\psi} \gamma^{\mu} \partial_{\mu} \psi
 - \left( {\cal W}''^2 + {\cal W}' {\cal W}''' \right) \phi'^2 
- {\cal W}'' \bar{\psi} \psi \right\}.
\end{equation}
The equations of motion for the scalar and the fermion are
\begin{equation}
\partial^2\phi' + \left\{ {\cal W}''^2 + {\cal W}' {\cal W}'''\right\} \phi' =0,
\end{equation}
and
\begin{equation}
\imath \gamma^{m} \partial_{m} \psi - {\cal W}'' \psi =0.
\end{equation}

The scalar zero mode related to the broken translational symmetry in the
$\hat{y}$ direction is
\begin{equation}
\phi'(x,y,t)=\frac{d\phi_w}{dy}(y) A(x,t),
\end{equation}
where $A(x,t)$ satisfies the equation of motion of a massless real scalar,
$\partial^2 A=0$, in $d=1+1$ dimensions.
By acting with a supersymmetry generator on $\phi_w$ we find
\begin{equation}
\delta \psi = - \alpha \frac{d\phi_w}{dy}(1-\alpha\imath\gamma^2)\zeta.
\end{equation}
For the unbroken generator $\zeta=1/2 (1+\alpha\imath\gamma^2)\zeta$, whereas
for the broken generator $\zeta=1/2 (1-\alpha\imath\gamma^2)\zeta$. 
The fermionic zero mode related to the broken supersymmetry generator is
\begin{equation}
\psi(x,y,t)=\frac{d\phi_w}{dy}(y) \zeta(x,t),
\end{equation}
where $\zeta$ satisfies the massless Dirac equation in $d=1+1$ dimensions
$\imath \gamma^{a}\partial_a\zeta=0$, with $a=0,1$, satisfies the Majorana constraint and
in addition the constraint
$\zeta=1/2 (1-\alpha\imath\gamma^2)\zeta$, that is, $\zeta$ is chiral. 
Consequently, 
the massless fermion on the wall is either left or right moving 
depending on the sign of $\alpha$.

\subsubsection*{${\cal N}=2$ SUSY}

The minimal model with ${\cal N}=2$ supersymmetry in $d=2+1$ dimensions contains two
${\cal N}=1$ supermultiplets. The Lagrangian in components is
\begin{eqnarray}
\cal{L} & = & \frac{1}{2} \left\{ \partial_m \phi_i \partial^m \phi_i + 
\imath \bar{\psi}_i \gamma^{m} \partial_{m} \psi_i + F_i F_i \right. \nonumber \\
 & & \left. +2 F_i \frac{\partial{\cal W}}{\partial\phi_i} 
-\frac{\partial^2{\cal W}}{\partial\phi_i\partial\phi_j} \bar{\psi}_i \psi_j \right\},
\end{eqnarray}
where the indices $i$, $j$ are $1,2$ and the superpotential is harmonic, so that
\begin{equation}
\frac{\partial^2{\cal W}}{\partial\phi_1^2} + \frac{\partial^2{\cal W}}
{\partial\phi_2^2}=0.
\end{equation}
An ${\cal N}=2$ model in $d=2+1$ dimensions can be obtained from an ${\cal N}=1$ model with one chiral
superfield in $d=3+1$ dimensions. In this procedure, the fields are assumed to be
independent of one of the spatial coordinates. The superpotential in the lower
dimensional theory is obtained from the superpotential in the higher dimensional
theory by
\begin{equation}
{\cal W}(\phi_1,\phi_2) = 2 {\rm Re}\left[W(\frac{1}{\sqrt{2}}
(\phi_1+\imath \phi_2))\right].
\end{equation}
The lower dimensional superpotential obtained in this way is automatically harmonic,
because the higher dimensional superpotential is holomorphic.
The $N=2$ model in $d=2+1$ dimensions has, apart from the regular supersymmetry
transformations of $N=1$ models,
\begin{equation}
\begin{array}{ll}
\delta \phi_1 =  \bar{\zeta} \psi_1\, , & \delta \phi_2 =  \bar{\zeta} \psi_2\, , \\
\delta \psi_1 =  - \imath \partial_{m} \phi_1 \gamma^{m} \zeta + F_1\zeta\, , &
\delta \psi_2 =  -\imath \partial_{m} \phi_2 \gamma^{m} \zeta + F_2\zeta\, , \\
\delta F_1 = - \imath \bar{\zeta}\gamma^{m}\partial_{m} \psi_1\, , &
\delta F_2 = - \imath \bar{\zeta}\gamma^{m}\partial_{m} \psi_2,
\end{array}
\end{equation}
additional supersymmetry transformations that take the form:
\begin{equation}
\begin{array}{ll}
\delta \phi_1 = - \bar{\eta} \psi_2\, , & \delta \phi_2 =  \bar{\eta} \psi_1\, , \\
\delta \psi_1 = - \imath \partial_{\mu} \phi_2 \gamma^{\mu} \eta - F_2\eta\, , &
\delta \psi_2 =  \imath \partial_{\mu} \phi_1 \gamma^{\mu} \eta + F_1\eta\, , \\
\delta F_1 = -\imath \bar{\eta}\gamma^{\mu}\partial_{\mu} \psi_2\, , &
\delta F_2 = \imath \bar{\eta}\gamma^{\mu}\partial_{\mu} \psi_1.
\end{array}
\end{equation}
From the perspective of dimensional reduction, these extra 
transformations arise because the underlying $d=3+1$ dimensional 
theory has four supersymmetry generators.

The auxiliary fields $F_i$ can be eliminated to give the Lagrangian 
\begin{equation}
\label{Lag1+1}
{\cal L} = \frac{1}{2} \left\{ \partial_{m}\phi_i \partial^{m}\phi_i + \imath
\bar{\psi}_i \gamma^m \partial_m \psi_i - \frac{\partial {\cal W}}{\partial \phi_i}
\frac{\partial {\cal W}}{\partial \phi_i} - \frac{\partial^2 {\cal W}}{\partial \phi_i
\partial \phi_j} \bar{\psi}_i \psi_j \right\}.
\end{equation}
The BPS equation is
\begin{equation}
\frac{d\phi_i}{dy} = \alpha \frac{\partial {\cal W}}{\partial \phi_i},
\end{equation}
with $\alpha=\pm 1$. 

Expanding the Lagrangian of Eq.\,(\ref{Lag1+1}) about the 
solution $\phi_i^w$ up to quadratic fluctuations, we obtain
\begin{equation}
{\cal L} = \frac{1}{2} \left\{ \partial_m \phi_i'\partial^m \phi_i'
+ \imath \bar{\psi}_i \gamma^{m} \partial_m \psi_i 
- {\cal W}_i {\cal W}_{ikl} \phi_k' \phi_l' - {\cal W}_{ik} {\cal W}_{il} \phi_k' \phi_l'
- {\cal W}_{ij} \bar{\psi}_i \psi_j \right\},
\end{equation}
where the subscripts on $\cal W$ indicate derivatives with respect to $\phi_i$ evaluated
at $\phi_i=\phi_i^w$.

In the domain wall background, the scalar equation of motion is 
\begin{equation}
\partial_m \partial^m \phi_n' + \left\{ {\cal W}_i {\cal W}_{inl} + {\cal W}_{in}
{\cal W}_{il} \right\} \phi_l'=0,
\end{equation}
and fermions satisfy
\begin{equation}
\imath \gamma^{m} \partial_{m} \psi_i - {\cal W}_{ij} \psi_j =0.
\end{equation}
There is a massless scalar, $A(x,t)$, on the wall which is related to the
broken translational symmetry,
\begin{equation}
\phi_i'(x,y,t)= \frac{d\phi_i^w}{dy}(y) A(x,t)\, ,
\end{equation}
and a massless fermion, $\zeta(x,t)$, (either right or left moving, depending 
on the
sign of $\alpha$), due to the broken 
regular supersymmetry generator,
\begin{equation}
\psi_i= \frac{d\phi_i^w}{dy} \zeta(x,t),
\end{equation}
where $\zeta$ is a (real) two component Majorana spinor satisfying 
the $1+1$ dimensional massless Dirac equation
and the constraint $\zeta=1/2 (1-\imath \alpha \gamma^2) \zeta$.
In addition, one of the extra SUSY generators is broken,
which results in another massless fermion on the wall,
\begin{eqnarray}
\psi_1(x,y,t) & = & \frac{d\phi_2^w}{dy} \eta(x,t), \nonumber \\
\psi_2(x,y,t) & = & - \frac{d\phi_1^w}{dy} \eta(x,t), 
\end{eqnarray}
where $\eta$ is another (real) two component Majorana spinor satisfying 
the $1+1$ dimensional Dirac equation
and the constraint $\zeta=1/2 (1+\imath \alpha \gamma^2) \zeta$. We therefore
find in the case of the BPS wall in the $d=2+1$, ${\cal N}=2$ case a massless
scalar, a left moving massless fermion and a right moving massless
fermion, which together form an ${\cal N}=1$ multiplet in the lower dimensional
theory.

\bibliographystyle{prsty}

\end{document}